\documentclass{eas}
\usepackage{graphicx}
\usepackage{natbib}
\newcommand\apj{{ApJ}}%
\newcommand\aap{{A\&A}}%
\newcommand\mnras{{MNRAS}}%
\newcommand\pasa{{PASA}}%
\begin{document}

\title{Stellar Motion around Spiral Arms: {\it Gaia} Mock Data} 
\runningtitle{Kawata \etal: Stellar Motion around Spiral Arms}
\author{Daisuke Kawata}\address{Mullard Space Science Laboratory, University College London. Dorking, Surrey, RH5 6NT, UK}
\author{Jason A.S. Hunt}\sameaddress{1}
\author{Robert J.J. Grand}\address{Heidelberger Institut f{\"u}r Theoretische Studien, Schloss-Wolfsbrunnenweg 35, 69118 Heidelberg, Germany}\secondaddress{Zentrum f{\"u}r Astronomie der Universit{\"a}t Heidelberg, Astronomisches Recheninstitut, M{\"o}nchhofstr. 12-14, 69120 Heidelberg, Germany}
\author{Arnaud Siebert}\address{Observatoire Astronomique, Universit\'e de Strasbourg, CNRS, 11 rue de l'universit\'e 67000 Strasbourg, France}
\author{Stefano Pasetto}\sameaddress{1}
\author{Mark Cropper}\sameaddress{1}
\begin{abstract}
We compare the stellar motion around a spiral arm created in two different scenarios, transient/co-rotating spiral arms and density-wave-like spiral arms. We generate {\it Gaia} mock data from snapshots of the simulations following these two scenarios using our stellar population code, {\tt SNAPDRAGONS}, which takes into account dust extinction and the expected {\it Gaia} errors.
We compare the observed rotation velocity around a spiral arm similar in position to the Perseus arm, and 
find that there is a clear difference in the velocity features around the spiral arm between the co-rotating spiral arm and the density-wave-like spiral arm. Our result demonstrates that the volume and accuracy of the {\it Gaia} data are sufficient to clearly distinguish these two scenarios of the spiral arms. 
\end{abstract}
\maketitle
\section{Introduction}

Since the 1960s, the most widely accepted explanation of the spiral arms has been that these features are wave structures, i.e. the density-wave scenario, and rotating rigidly (a constant pattern speed) with a different speed from those of the stars \citep{ls64}. However, recent observations of external galaxies show evidence against long-lived spiral arms with a constant pattern speed. For example, using the modified Tremaine-Weinberg method, \citet{mrm06} suggested a radially decreasing pattern speed for the spiral arms in NGC 1068. Also, by combining H$\alpha$ imaging and $Swift/UVOT$ Near-Ultraviolet (NUV) data, \citet{fckph12} distinguished the regions with ongoing star formation and the regions with star formation a few hundred million years ago in the grand-design spiral galaxy, M100. Contrary to the expectation from the density-wave theory, no offset was found between these two regions. The same conclusion was reached in \citet{frdlw11} for different galaxies. 

The problem is also evident on the theoretical side by using N-body simulations. Even with the recent high-resolution numerical simulations not a single N-body simulation is able to reproduce a long-standing spiral arm feature as suggested by the density-wave scenario \citep[e.g][]{jas11,db14}. \citet{gkc12a,gkc12b} demonstrated that the spiral arm was rotating with the same speed as the stars, i.e. co-rotating, at every radii, and therefore winding \citep[see also][]{rvfrv13}. Still, in each snapshot, the spiral arms are always apparent, and the spiral arms are constantly forming and disrupting, i.e. recurrent, with a lifetime of about 100 Myr. Although co-rotating spiral arms lead to the winding-dilemma, \citet{gkc13} demonstrated that the spiral arms were disrupted before they wound up completely, and the pitch angle of the spiral arms correlated with the shear rate of the disc, as observed \citep[e.g.][]{sbbh06}. Interestingly, the winding nature of the spiral arms seen in N-body simulations can naturally explain the observed scatter in the correlation between the pitch angle and the shear rate \citep[see][for more thorough discussion]{gkc13}.

The Milky Way is a (barred) spiral galaxy which we can observe in great detail. The influence of the spiral arms on the stellar motion has also been measured and compared with the models \citep[e.g.][]{fft01,sfbbf12,fsf14}. The European Space Agency (ESA) {\it Gaia} mission will produce accurate positions and velocities for over a billion stars in the Milky Way, and the majority of them are the disk stars.  The {\it Gaia} data will provide the opportunity for astronomers to observe in great detail how the stars are moving around the spiral arms in a spiral galaxy, and should provide decisive constraints on the different spiral arm scenarios.

\section{Stellar motion around the transient/co-rotating spiral arms}

  The density enhancement of the co-rotating spiral arms causes an efficient radial migration for both stars and gas \citep{gkc15}, because the stars cannot pass or be passed by the spiral arm. For example, the stars behind the spiral arm are accelerated by the potential of the spiral arm, and their guiding centre moves outwards. Because they are rotating with the same speed as the spiral arm, they stay behind the spiral arm and keep being accelerated. \citet{gkc14} showed that the stars migrating outward are always behind the spiral arm and at apo-centre phase, i.e. rotating slower than the circular velocity at that radius, while the stars migrating inwards are in the front of the spiral arm and at peri-centre phase. \citet{hkgmpc15} generated star samples from an N-body/smoothed particle hydrodynamics simulation of a Milky Way-sized barred disk galaxy in \citet{khgpc14} using {\tt SNAPDRAGONS} (Stellar Numbers And Parameters Determined Routinely And Generated Observing N-body Systems). {\tt SNAPDRAGONS} generates the stellar samples taking into account the metallicity and age of a star particle, and produces the catalogue of stars which are observed from the assumed observer position, taking into account an empirical 3D extinction \citep{sbhjb11} and the expected {\it Gaia} errors \citep[using the updated version of][]{rgfaaa15}. \citet{hkgmpc15} showed that {\it Gaia} has sufficient accuracy to detect the difference in the velocity distribution behind and in the front of the Perseus arm.

\begin{figure}
\centering
\includegraphics[width=\hsize]{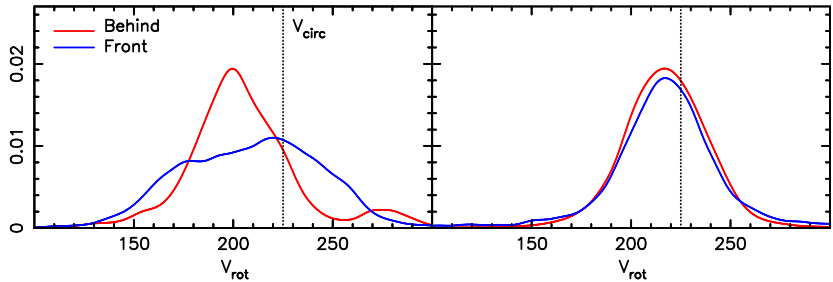}
\caption{Galactic rotation velocity distribution of the stars ($V<16$ mag) generated with {\tt SNAPDRAGONS} from the particles in the Galactic longitude of $85<l<95$ deg and latitude of $-5<b<5$ deg in an N-body simulation which shows the transient/co-rotating spiral arm (Left) and a test particle simulation of \citet{fsf14} where the density-wave-like rigidly rotating spiral arm is assumed (Right). Red (blue) line shows the kernel density estimation of the rotation velocity distribution of stars behind (in the front of) the spiral arm which is about 4 kpc away from the observer. The vertical dotted line indicates the circular velocity. Note that the extinction and the expected end-of-mission {\it Gaia} errors are included. }
\label{vrothis-fig}
\end{figure}

\section{Transient/co-rotating spiral arms vs. density-wave-like spiral arms}

To compare with the stellar motion around the spiral arm in the N-body simulation in \citet{hkgmpc15}, we also made mock {\it Gaia} data using the 3D test particle simulation data with rigidly rotating, density-wave-like, spiral arms in \citet{fsf14}. Fig.~\ref{vrothis-fig} shows the rotation velocity distribution of the stars behind and in the front of the arm at the Galactic longitude, $l=90$ deg, for the N-body simulation (left) and the test particle simulation (right). The position of the observer is set so that there is a spiral arm at the similar distance to the Perseus arm at $l=90$ deg. We consider only the stars with $V<16$ mag for which {\it Gaia} RVS will produce the accurate line-of-sight velocity and the rotation velocities will be measured with reasonably high accuracy.

In the left panel of Fig.~\ref{vrothis-fig}, the peak rotation velocity for stars behind the transient/co-rotating spiral arm is slower than the circular velocity, and the stars in the front of the spiral arm shows a broader distribution. No such difference is observed in the right panel of the results of the density-wave-like spiral arm. This demonstrates that the rotation velocity distribution behind the transient/co-rotating spiral arm in the N-body simulation is different from that in front of the spiral arm, and this is clearly different from the density-wave-like spiral arm in the test particle simulation. This result also demonstrates that the accuracy of {\it Gaia} data is superb, and {\it Gaia} will provide enough information to trace the stellar motion on both sides of the Perseus spiral arm. In transient/co-rotating spiral arms in N-body simulations, the difference in the velocity distribution is expected to be observed at every radius \citep{khgpc14}. We can therefore carry out similar analysis to Fig.~\ref{vrothis-fig} at different radii, i.e. different longitudes, and directly compare with the {\it Gaia} data once they become publicly available. The {\it Gaia} data should therefore be able to distinguish between these two different scenarios of the spiral arms.




\end{document}